\documentclass[MSNbibl,nameyear,dvips]{arxstspdf}
\usepackage{flushend}
\usepackage{stfloats}

\volume{26}
\issue{2}
\pubyear{2011}
\firstpage{262}
\lastpage{265}
\doi{10.1214/11-STS346C}
\referstodoi{10.1214/10-STS346}

\makeatletter
\let\epsilon\varepsilon
\makeatother

\begin{document}
\begin{frontmatter}

\title{Discussion of
``Impact of Frequentist and Bayesian Methods on Survey Sampling
Practice:
A Selective Appraisal'' by J.~N.~K.~Rao}
\runtitle{Discussion}
\pdftitle{Discussion of Impact of Frequentist and Bayesian Methods on
Survey Sampling Practice: A Selective Appraisal by J. N. K. Rao}

\begin{aug}
\author[a]{\fnms{Eric} \snm{Slud}\corref{}\ead[label=e1]{evs@math.umd.edu}}
\runauthor{E. Slud}

\affiliation{University of Maryland}

\address[a]{Eric Slud is Professor, Statistics Program, Department of Mathematics, University of Maryland, College Park, Maryland 20742, USA
\printead{e1}.}

\end{aug}

% ABSTRACT

% KEYWORDS

\end{frontmatter}

I would like to congratulate Professor Rao on having produced
an overview of survey methodology which is at the same time a
broad-ranging prospectus of current research and also an impressive
retrospective from a modern viewpoint of the early historical
developments. He shows us in broad terms where the various approaches
to survey methodology have been successful and where they cannot
quite be relied upon without further development.

Most of the paper is not specifically directed at contrasting the
Bayesian and frequentist viewpoints. The most important distinctions
for Rao seem to be between model-dependent and design-based methods,
and Bayes methods are faulted in Rao's chosen terrain of ``the
large-scale production of official statistics from complex surveys''
primarily for using models where models are not absolutely necessary.
He takes for granted that models will be used in adjusting for
nonresponse, in his formulation largely through calibration, and
in small area estimation. The faults he finds with unnecessarily
model-dependent survey estimation methods are:
\begin{itemize}
\item design-inconsistency (of model-based BLUP under misspecified
models, and in other examples, in Section~3.2);
\item requiring different sets of predictor variables for different
attributes of interest (in Section~3.3);
\end{itemize}
and in Section~4.2, in relation to the nonparametric Bayesian and
pseudo-Bayesian methods relying heavily on exchangeability, for their
\begin{itemize}
\item lack of generalizability to complex survey designs with
clustering and unequal probability \mbox{weighting}.
\end{itemize}
Like many authors in survey sampling, Rao faults model-based analyses
because of possible model misspecification. This discussion highlights
aspects and consequences of model misspecification under the headings
of Rao's paper.

%s1 ###
\section{Model Misspecification in Linear Regression and
Calibration}\label{sec1}

In Section 3.1 of his paper, Rao considers the
behavior of a calibration estimator (of a population total) when
the calibration constraints involve some but not all of the predictor
variables entering a true superpopulation model.
The context is a superpopulation in which the regression model
%e1 ###
\begin{equation}\label{Mod1}
Y_i = \beta' X_i + \gamma' Z_i + \epsilon_i
\end{equation}
holds for all units $i$ in the frame $  \mathcal{U}$, with auxiliary
variables $ X_i, Z_i $ known for all population units, and where it is
desired to estimate the total $ t_Y = \sum_{i \in \mathcal{U}}   Y_i $
based on a probability sample of units \mbox{$ i \in \mathcal{S} $} with first-order
inclusion weights $ d_i=1/\pi_i$.  [In Rao's example, the weights $ d_i $
are all equal, $ X_i=(1,x_i)'$,  and $ Z_i = x_i^2$,  for a scalar auxiliary
varia\-ble $ x_i$.] A~calibration estimator of $ t_Y $ might be based on the
varia\-bles~$X_i$ alone, that is, on $ \sum_{i \in \mathcal{S}}   w_i   Y_i $ where
the  modified weights $ w_i $ are determined by minimizing\break
$ \sum_{i\in \mathcal{S}}   (w_i-d_i)^2/d_i $ subject to the constraints\break
$ \sum_{i\in \mathcal{S}}   w_i   X_i   =   \sum_{i\in \mathcal{U}}   X_i$.
As described by Rao, it turns out that this calibration estimator is
equivalent to the generalized regression (GREG) estimator based on the weights
$ d_i $ and the predictor variable $ X_i$.  In the setting with constant $d_i$,
this estimator would be the unweighted model-based regression estimator
based on predictor $ X_i$.

As Rao suggests, calibration might be based on a~subset of the
appropriate predictor variables when the same universal calibration
constraints are used\vadjust{\eject} over many different choices of response
variables. In the context (\ref{Mod1}) above, there are three ways in which
this calibration estimator based on variables~$ X_i $\break might be
inadequate. First, the weights $ d_i $ used in the estimator might not
be the correct ones: for example, when the unweighted regression
estimator is used but the design weights are not constant, this is a
familiar kind of wrong-model inconsistency that arises in Section 3.2.
Second, the calibration totals $ \sum_{i\in \mathcal{U}}   X_i $ fixed
in defining the estimator might not be correct: this may be viewed as
a failure of the frame-coverage model. [A superpopulation-based
treatment of linear calibration with inaccurate totals is given in
Slud and Thibaudeau (\citeyear{SluThi}), Proposition 1, in a~more general setting also
involving nonresponse adjustment and weight-compression.] Third, as
mentioned in Rao's paper with reference to Rao, Jocelyn and Hidiroglou (\citeyear{RAOJOCHID03}), the
coverage of the confidence intervals for $ t_Y $ based on this
calibration estimator might not be close to nominal in moderate
samples.  The first two of these three cases represent actual design
inconsistency.  However, if the weights and calibration totals are
correct, then the calibration estimator based on $ X_i $ is still a
model-assisted GREG estimator and therefore design-consistent under
general conditions, but the problematic coverage of its confidence
intervals seems to be due to slow convergence to the limiting normal
asymptotic distribution, which Rao, Jocelyn and Hidiroglou (\citeyear{RAOJOCHID03}) found to be related
to skewness of the residuals from the incorrect linear regression
model of $Y_i$ on~$X_i$ when (\ref{Mod1}) holds with nonzero $\gamma$. This
failure of mode\-rate-sample coverage of confidence intervals due to
slow distributional convergence is more subtle than
design-inconsistency, but may still be important in practice in
surveys where regressions are done separately in each stratum, since
the whole sample might be large while the individual strata might all
have moderate sample size.

%s2 ###
\section{Diagnostics in Small Area Estimation}\label{sec2}

One survey-sampling task where all practitioners would agree on the
necessity of explicit models is Small Area Estimation. When survey
estimates are required for small domains where little or no sample
is available, models perform a function of driving direct estimates
toward covariate-defined predictors, providing extrapolated
estimates in domains where there is no sample and shrinking direct
estimates for covariate-defined similar domains together. The most
convenient small area estimation models, whe\-ther hierarchical
Bayes or generalized-linear with aggregate-level random effects,
have the same form for all domains in the frame population. For any
specific proposed model, this is an assumption that requires checking
and may prove crucial to the quality of small area estimates
or predictions. Yet there is remarkably little work on goodness-of-fit
checking in small area models, and hardly any mention of the topic
in the present paper, due in part to Rao's focus in Section 5 on
Hierarchical Bayes methods.

Goodness-of-fit and model-checking methods have been studied in the
survey literature, with important contributions by Rao himself.
Chi-squared tests based on survey cross-classifications were studied
in a series of papers leading up to Rao and Scott (\citeyear{RaoSco84}),
and are widely cited but perhaps not much used in model-checking. A different
chi-squared test, based on estimated cell-frequencies in multi-way tables
and suited to small area models, was given by Jiang, Lahiri and Wu (\citeyear{JiaLahWu01}),
work which was extended to tests for mixed linear model diagnostics studied in Jiang (\citeyear{Jia01}), again in a form which could be used in assessing
the fit of a small area model. In a different direction, the paper of Eltinge
and Yansaneh (\citeyear{EltYan97}) is unusual in providing diagnostics for nonresponse
adjustment cells in surveys. Apart from these papers, diagnostics are often
borrowed from parametric nonsurvey statistics in individual survey applications.

The Census Bureau's Small Area Income and Po\-ver-ty Estimates (SAIPE)
program, mentioned by Rao as a source of examples for small area
methodology, has provided an extensive test-bed for small area model-checking techniques (Citro and Kalton, \citeyear{CitKal}). As described in Rao
[(\citeyear{Rao03}), Chapter~7] and Citro and Kalton (\citeyear{CitKal}), the county-level log-count
model for poor children had the Fay--Herriot form
%e2 ###
\begin{eqnarray}\label{Mod2}
y_i = x_i'\beta + u_i + e_i, \nonumber\\[-8pt]\\[-8pt]
\eqntext{u_i \sim \mathcal{N}(0,\sigma^2),
   \ e_i \sim \mathcal{N}(0,v_e/n_i),}
\end{eqnarray}
where $y_i$ is the direct-estimated log-count of poor children in
county $i$,  $x_i$ is  a vector of covariate predictors, $ n_i $ is
the number of sampled households, $ u_i $ is the county-level random effect,
and $ e_i$  are random survey errors with variances assumed known. Because
roughly 20\% of sampled counties, with positive $ n_i$,  yielded no poor
children and therefore would have provided direct estimates of $ 0 $ poor
children, the logarithms of these estimates are undefined and those counties
were dropped from the model-fitting analysis. Despite the very effective
small area predictions generated by fitting unknown parameters $\beta$
to the set of sampled counties with well-defined $y_i$, it remains
questionable whether that fitted model (\ref{Mod2}) should be used to
predict numbers of poor children in counties where no poor children
were seen. This is an issue of model specification, which has been
studied for a number of years (Slud, \citeyear{SluN1}, \citeyear{SluN2}) and for which
diagnostics have now been developed in Slud and Maiti (\citeyear{SluMai}) by
regarding the dropped counties as having been left-censored (or
left-truncated) because they are dropped when the count of sampled
poor children is below a threshold. These diagnostics seem to show
that the model (\ref{Mod2}) adequately describes the counties with
well-defined $y_i$, but that the same model cannot adequately predict
in which counties there would be any poor children in a sample. The
upshot is that no model is yet known which can account for counts of
sampled poor children in all counties.

%s3 ###
\section{Specification of Multilevel Survey Analyses}\label{sec3}

The kind of model-checking described in the previous paragraph is
important because, while it is common for survey data sets (including
aggregated area-level data sets used in small area modeling)
to be highly cross-classified by covariates as well as unit response
versus nonresponse, there is no guarantee that a single
model can account well for all portions of the cross-classified
population. Such survey data naturally suggest multilevel models,
but models which differ in form on different subsets of the population
would lead to complicated interaction terms and random effects.

Rao's paper treats multilevel modeling in a frequentist
design-based setting in Section 3.3, under the general heading of
estimation in complex surveys; yet when discussing unified models in a
small area context, he accepts the value of hierarchical-Bayes models.
Why is that? In general complex surveys, it seems likely that
simultaneous hierarchical-Bayes (HB) models could be formulated for
unit nonresponse, frame coverage errors, and survey responses. If
reasonable rules could be developed for defining prior parameters,
then a Bayesian analysis is not on its face less theoretically
acceptable than a~complicated weight-adjustment procedure.  But
perhaps one serious objection is that each response variable would
require its own Bayesian mo\-del. Is the greater value of HB models for
small area prediction due to the acceptability in that context of a~%
separate model for each survey response variable?

In the small area context, my own view is that hierarchical-Bayes
models with objective priors---or priors chosen by the matching
strategies discussed in Section \ref{sec4}---might very well serve the
smoothing function of shrinking direct estimators from similar
areas toward one another. But I feel much less comfortable
with this class of models being used to extrapolate small area
predictions to areas with very small or zero sample sizes.

A difficulty with multilevel models, for both frequentists and
Bayesians, is that different hierarchical error structures can
sometimes be almost impossible to distinguish with useful power for
moderately large sample sizes, as may be revealed by
information-matrix calculations. Nevertheless, there are
data sets where (generalized) likelihood ratio testing for the
presence of certain error structure components can be rather
decisive. In a spatial small area problem, Opsomer et al. (\citeyear{Opsetal08})
modeled the alkalinity of lakes in a survey of lakes in terms
of elevation and radial P-spline basis functions in spatial
coordinates, with the spline-term coefficients as random effects.
In addition, independent random effects for slightly aggregated
geographic units were considered and found to be important after
likelihood ratio testing. It will not always be possible to reach
such firm conclusions, and this kind of model-comparison may be
hard to reproduce in a Bayesian framework.

%s4 ###
\section{Miscellaneous Comments}\label{sec4}

All of us, frequentists and Bayesians, are tied to models in the
sense that statistical theory generally has very little to say
about the validity of likelihood-based inferences when the
parametric model family does not contain the model actually
governing the data.

For sample survey data, frequentists have always found it
difficult to say what is an appropriate likelihood. [However, Rao's
paper mentions in Section~5 fascinating work in Wu and Rao (\citeyear{WuRao06}), Rao and
Wu (\citeyear{RAOWU10}), attempting to interpret empirical-likelihood survey methods
as a Bayesian nonparametric survey likelihood.] A design-based view of
finite-population sampling forces us to view the ensemble of survey
attributes as nuisance parameters, about which we are entitled to
assume only a sort of large-superpopula\-tion stability.  A frequentist
approach to the high nuisance-parameter dimension is to base
inferences on estimating equations, which is how Rao presents in
Section 3.3 the ``model-assisted'' pseudo-likelihood method of
estimating frame-population descriptive parameters, such as regression
coefficients via GREG, and such as the multilevel variance-component
parameters that are the target of multilevel survey estimation. As far
as I can tell, this approach has no Bayesian counterpart, so the
survey analyst who wants the protection of correct estimation for
virtually any superpopulation configuration of survey attributes has
little recourse but to follow design-based theory. That seems to be
the essence of the argument in favor of design-based survey methods
when models are not absolutely necessary because of missing data.

Weight adjustment for calibration and model-ba\-sed nonresponse
adjustment can also be viewed as estimating equation methods.
Like other such methods, weight-adjustments rely for their validity
on correctness of at least some model assumptions: as Rao mentions,
the most we can hope for in this enterprise is a kind of ``double
robustness'' in which design-consistency for the weighted survey
estimator obtains when either the model used for nonresponse adjustment
or a population-wide regression-type model is correct. See Kang and
Shafer (\citeyear{KanSch07}) for related exposition of the double-robustness concept,
and Slud and Thibaudeau (\citeyear{SluThi}) for analogous results on a further
development of the optimization-based weight-adjustment method of
Deville and\break S\"{a}rndal (\citeyear{DevSar92}) to cover simultaneous weight-adjust\-ment
for nonresponse, calibration and weight-com\-pression.

%% One final remark related to model misspecification mirrors ideas
%% told to me by Several Bayesian survey analysts at the 2008 Workshop
%% where the papers in this Special Issue were first presented.
Survey estimation is often an exercise in prediction, and it is known
in many statistical problems that excellent predictions can be
provided through estimating models which are too simple to pass
good\-ness-of-fit checks. This observation has not yet been formulated
with mathematical care---no one knows how to characterize which
target parameters and which combinations of true and oversimplified
models could work in this way---but frequentists and Bayesians would
all benefit from a rigorous result of this type.

% imsref loaded by dianan, 2011-05-06 11:08:31
\vspace*{-4pt}

\end{document}